\documentclass[a4paper,fleqn,usenatbib]{mnras}

\usepackage{newtxtext,newtxmath}

\usepackage[T1]{fontenc}
\usepackage{ae,aecompl}

\usepackage{amsmath}
\usepackage{graphicx,longtable,mathrsfs,color,array}
\usepackage{epsfig,subfigure,placeins,float}
\usepackage[usenames,dvipsnames]{xcolor}
\usepackage{hyperref}
\usepackage{multirow}
\usepackage{todonotes}
\usepackage{caption}

\usepackage[normalem]{ulem}
\providecommand{\adsurl}[1]{\href{#1}{ADS}}

\newcommand*{\rom}[1]{\expandafter\@slowromancap\romannumeral #1@}

\newcommand{\ms}{M$_{\sun}$}

\newcommand{\mcrit}{$M_{\rm 500crit}$}

\newcommand{\mtot}{$M_{\rm tot}$}
\newcommand{\msim}{$M_{\rm 500,SIM}$}
\newcommand{\mhse}{$M_{\rm 500,HSE}$}
\newcommand{\mspec}{$M_{\rm 500,SPEC}$}
\newcommand{\mspecc}{$M_{\rm 500crit}^{\rm SPEC}$}
\newcommand{\msimc}{$M_{\rm 500crit}^{\rm SIM}$}
\newcommand{\mhsec}{$M_{\rm 500crit}^{\rm HSE}$}
\newcommand{\myx}{$M_{Y_{\rm X}}$}

\newcommand{\ysz}{$Y_{\rm SZ}$}

\newcommand{\yszsim}{$Y_{\rm SZ, 500,SIM}$}
\newcommand{\yszspec}{$Y_{\rm SZ, 500,SPEC}$}

\newcommand{\yx}{$Y_{\rm X}$}
\newcommand{\yxsim}{$Y_{{\rm X, SIM}}$}
\newcommand{\yxspec}{$Y_{{\rm X, SPEC}}$}

\newcommand{\rcrit}{$R_{\rm 500crit}$}
\newcommand{\rsim}{$R_{\rm 500,SIM}$}

\newcommand{\rspec}{$R_{\rm 500,SPEC}$}

\newcommand{\xp}{$X_{\rm pivot}$}

\newcommand{\yxm}{\mbox{$Y_{\rm X}$ -- $M_{\rm tot}$}}
\newcommand{\yszm}{\mbox{$Y_{\rm SZ}$ -- $M_{\rm tot}$}}

\newcommand{\tng}{\mbox{IllustrisTNG}}

\newcommand{\val}[3][0.000]{$#1^{+#2}_{-#3}$}

\makeatother

\title[Impact of X-ray mass estimates on $Y$\,--$M$ relation]{
Unifying Sunyaev-Zel'dovich and X-ray predictions from clusters to galaxy groups: 
the impact of X-ray mass estimates on the $Y$\,--\,$M$ scaling relation}

\author[A. R. Pop et al.]{Ana-Roxana Pop,$^{1}$\thanks{E-mail: ana-roxana.pop@cfa.harvard.edu}
Lars Hernquist,$^{1}$
Daisuke Nagai,$^{2}$
Rahul Kannan,$^{1}$ 
\newauthor
Rainer Weinberger,$^{3}$
Volker Springel,$^{4}$ 
Mark Vogelsberger,$^{5}$
Dylan Nelson,$^{6}$
\newauthor
R\"{u}diger Pakmor,$^{4}$
Paul Torrey$^{7}$
\\
$^{1}$Center for Astrophysics $\rvert$ Harvard \& Smithsonian, 60 Garden Street, Cambridge, MA 02138, USA \\
$^{2}$Department of Physics, Yale University, New Haven, CT 06520, U.S.A. \\
$^{3}$Canadian Institute for Theoretical Astrophysics, 60 St. George Street, Toronto, ON M5S 3H8, Canada \\
$^{4}$Max-Planck-Institut f\"{u}r Astrophysik, Karl-Schwarzschild-Stra{\ss}e 1, D-85741 Garching bei M\"{u}nchen, Germany \\
$^{5}$Dept. of Physics, Kavli Institute for Astrophysics and Space Research, Massachusetts Institute of Technology, Cambridge, MA 02139, USA \\
$^{6}$Universit\"{a}t Heidelberg, Zentrum f\"{u}r Astronomie, Institut f\"{u}r theoretische Astrophysik, Albert-Ueberle-Str. 2, 69120 Heidelberg, Germany \\
$^{7}$Department of Physics, University of Florida, Gainesville, FL 32611, USA
}
\date{Accepted XXX. Received YYY; in original form ZZZ}

\pubyear{2022}

\begin{document}
\label{firstpage}
\pagerange{\pageref{firstpage}--\pageref{lastpage}}
\maketitle

\begin{abstract}
One of the main limitations in precision cluster cosmology arises from systematic errors and uncertainties in estimating cluster masses.
Using the Mock-X pipeline, we produce synthetic X-ray images and derive cluster and galaxy group X-ray properties for a sample of over 30,000 simulated galaxy groups and clusters with \mcrit\, between $10^{12}$ and $2\times 10^{15}$\,\ms\ in IllustrisTNG. We explore the similarities and differences between \tng\, predictions of the Sunyaev-Zel'dovich and X-ray scaling relations with mass. 
We find a median hydrostatic mass bias $b = 0.125 \pm 0.003$ for \mcrit\,  $>10^{13}$\,\ms. 
The bias increases to $b = 0.17 \pm 0.004$ when masses are derived from synthetic X-ray observations. 
We  model how different underlying assumptions about the dependence of \yx\, on halo mass can generate biases in the observed \ysz\, -- \myx\, scaling relation. 
In particular, the simplifying assumption that \yxm\, is self-similar at all mass scales largely hides the break in \yszm\, and overestimates \ysz\, at galaxy and groups scales. 
We show that calibrating the \yx--mass proxy using a new model for a smoothly broken power law reproduces the true underlying \yszm\, scaling relation with high accuracy. Moreover, \myx\, estimates calibrated with this method lead to \ysz\, -- \myx\, predictions that are not biased by the presence of lower mass clusters or galaxy groups in the sample.
Finally, we show that our smoothly broken power law model provides a  robust way to derive the \yx--mass proxy, significantly reducing the level of mass bias for clusters, groups, and galaxies.
\end{abstract}


\section{Introduction}

Galaxy clusters are the largest gravitationally bound objects in the Universe and they serve as one of the most powerful cosmological probes. To date, X-ray and microwave observations of galaxy clusters have been used to provide a variety of cosmological constraints \citep{Vikhlinin2009a,Sievers2013,PlanckCollaboration2014b,Mantz2015,Bocquet2019}. Over the next decade, ongoing and upcoming X-ray and microwave surveys, such as \textit{eROSITA} \citep{Bulbul2021} and CMB-S4/HD \citep{Raghunathan2022}, will yield observations of more than $10^5$ groups and clusters, increasing current observational samples by orders of magnitude. With samples of this magnitude, precision cosmology will enter a new epoch in which systematic uncertainties dominate over statistical errors.

In order to take advantage of the full statistical power of ongoing and upcoming X-ray and SZ surveys, we need accurate and robust estimates of the total masses of haloes, \mtot, and calibration of the observable-mass scaling relations \citep[][for a recent review]{Pratt2019}. For example, the tight correlations in the \yxm\, and \yszm\ scaling relations have led \yx\, and \ysz\, to be proclaimed as the most robust X-ray \citep{Kravtsov2006} and SZ  \citep[e.g.,][]{Nagai2006,Battaglia2012} mass estimators. Thus, due to the very low scatter in the \yxm\, and \yszm\, scaling relations, \yx\, and \ysz\, are the most widely adopted X-ray and SZ mass proxies \citep[e.g.,][]{Vikhlinin2009a,PlanckCollaboration2014b}.

As observations begin to probe the regime of low mass clusters and groups, the departure from the self-similar $Y-M$ scaling relation \citep[e.g.,][]{LeBrun2014,Planelles2014,McCarthy2017,Barnes2017b,Barnes2017a,Henden2018,Lim2021,Yang2022} and deviations from hydrostatic equilibrium \citep[e.g.,][]{Rasia2006,Nagai2007a,Biffi2016,Barnes2021} become more pronounced. Any cosmological study that makes simplistic assumptions about the scaling relations and their evolution can lead to significant biases in the mass estimation of clusters and groups.

In a companion paper (Pop et al. (2022), denoted as \mbox{Paper I} hereafter), we found a significant deviation from self-similarity for low mass clusters and galaxy groups across all X-ray and SZ observables -- mass relations in IllustrisTNG. Modelling the break from self-similarity necessitates adopting a model with sufficient flexibility to characterize the smooth transition between the power law observed for the highest mass clusters and the trends predicted for galaxy groups. In this paper, we investigate the behaviors of the robust mass proxies \yx\ and \ysz\ and develop a method for estimating the masses of galaxies, groups, and clusters using ongoing and upcoming X-ray and SZ surveys. 

This paper is organized as follows. In Section~\ref{sec:methods_yszmyx}, we present the methods used in this paper, including the sample of \tng\, haloes employed, the Mock X-ray pipeline that we developed in order to generate synthetic X-ray observations, and the method we use to derive hydrostatic and spectroscopic mass estimates. We present results for the hydrostatic and spectroscopic mass bias of \tng\, groups and clusters in Section~\ref{sec:massbias_yxysz}. In Section~\ref{sec:yxyszmassevolution}, we investigate the relationship between the \yszm\, and \yxm\, scaling relations as a function of halo mass, including the location of the break from self-similarity. We show how different assumptions about the underlying \yxm\, scaling relation can introduce significant biases in X-ray and SZ based cluster mass estimates in galaxies, groups, and low-mass clusters.  We present an analysis of those biases and show that calibrations utilizing a smoothly broken power law will reproduce the true \yszm\, relation with high accuracy. In Section~\ref{sec:unifying}, we present an overview of the mass biases introduced by different models and we show that the smoothly broken power law model provides a robust way to calibrate X-ray masses from \yx\, measurements, for a wide range of mass scales, \mcrit\,$\in [10^{12} - 2\times 10^{15}]$\,\ms. We conclude by summarizing our results in Section~\ref{sec:conclusions_letter}.

\section{Methods}
\label{sec:methods_yszmyx}

\subsection{Sample selection}
\label{subsec:sample_yszmyx}

In this work, we select and analyze all central objects (with \mcrit \, $\geq 10^{12}$\,\ms) in the \mbox{$z=0$} snapshot of the TNG300 simulation of the IllustrisTNG suite \citep{Nelson2018,Pillepich2018b,Springel2018,Naiman2018,Marinacci2018}, which is the successor of the Illustris simulation \citep{Vogelsberger2014b,Vogelsberger2014a,Genel2014}. Our final sample includes over 30,000 haloes in total. Over 2,500 of these haloes have \mcrit \, $\geq 10^{13}$\,\ms\,, and there are more than 150 clusters in our sample (\mcrit \, $\geq 10^{14}$\,\ms). In IllustrisTNG, haloes are identified through a friends-of-friends (FoF) algorithm with linking length $b=0.2$. 
Particle types such as gas, stars, and black holes are assigned to the same halo as their closest dark matter particle.
Galaxies / ``subhaloes" are identified by running the \textsc{subfind} \citep{Springel2001, Dolag2009} 
halo finder to identify gravitationally bound substructures that include all particle types. 

\subsection{X-ray pipeline}
\label{subsec:xraypipeline_yszmyx}

The multi-temperature structure of the intra-cluster medium can bias the observable properties of hot gas such as termperatures and X-ray luminosities \citep[e.g.,][]{Mazzotta2004, Rasia2005, Nagai2007a, Rasia2014}. Gas clumping and inhomogeneities can further affect the measurement \citep[e.g.,][]{Nagai2011, Zhuravleva2013, Khedekar2013, Vazza2013, Avestruz2014}. It is therefore critical to produce and analyze synthetic observational data from simulations using methods that closely match the pipelines used in observations. 

We generate mock X-ray observations for all the haloes in our sample using a modified version of the \textsc{Mock-X} pipeline introduced in \citet{Barnes2021}. 
Our method for computing synthetic X-ray observations mirrors previous approaches in this area 
 \citep[e.g.,][]{Gardini2004, Nagai2007a, Rasia2008, Heinz2009, Biffi2012, ZuHone2014, LeBrun2014, Henden2018}.

First, we begin by generating a rest-frame X-ray spectrum in the 0.5\,--\,10 keV band, using the density, temperature, and metallicity of every gas cell within a radius $1.5$\,\rcrit\, from the potential minimum of each halo.  We compute the X-ray spectrum using \textsc{apec} 
\citep[][]{Smith2001} with the \textsc{pyatomdb} module and atomic data from \textsc{atomdb} v3.0.9 \citep[see more details in][]{Foster2012}.
A particle's spectrum is the superposition of the individual spectra for each chemical element (H, He, C, N, O, Ne, Mg, Si, S, Ca, and Fe), scaled by the particle's elemental abundance. We exclude any gas with non-zero star formation rate, all cold gas cells with temperatures under \mbox{$10^5$\,K} and any gas cells that have a net cooling rate that is positive. The gas cells removed through this approach represent a sub-percent fraction of all gas cells within \rsim\, and they would not contribute significantly to the total X-ray surface brightness since most of their X-ray emission is outside the 0.5\,--\,10 keV energy band. 
This filtering removes uncertainties due to the imprecise thermal properties of gas cells with positive cooling or star-forming gas. This approach resembles the analysis of  X-ray observations, where compact sources with strong X-ray emission would be excised or simply unresolved. Previous simulation studies \citep[e.g.,][to name a few]{Nagai2007a, Henden2018, Barnes2021} have also utilized similar cuts.  Lastly, substructures are removed using the \textsc{subfind} halo finder, excluding all gas cells bound to galaxies other than the central object. 

For each halo in our sample, we produce mock X-ray spectra and model the spectroscopic temperature and X-ray luminosity of the object. We bin all gas cells within 0 -- 
1.5\,\rcrit\, of the halo center in 25 linearly spaced annuli. 
The energy range of 0.5\,--\,10 keV and energy resolution of 150 eV are set to match the \textit{Chandra} ACIS-I detector, and the resulting spectrum is convolved with the ACIS-I response matrix and the \textit{Chandra} effective area. 
In order to model the X-ray and SZ scaling relations down to galaxy and group scales (\mcrit\, $\geq 10^{12}$\,\ms), we use $10^6$ seconds-long exposures which are not limited by photon noise.
The resulting spectrum for each radial bin is modelled by a single-temperature and metallicity \textsc{apec} model in each radial annulus, and in this way we finally obtain the density, temperature and metallicity of the X-ray emitting gas as a function of radial distance.

\subsection{Hydrostatic and spectroscopic mass estimates}
\label{subsec:massbiasmethod}

In this subsection, we present our method for estimating the hydrostatic mass bias, as well as estimated masses derived from synthetic X-ray images. Several different methods for reconstructing mass profiles using X-ray data have been previously proposed \citep{Pointecouteau2005, Vikhlinin2006, Nagai2007a, Mahdavi2008, Nulsen2010, Sanders2018}, including reviews of the limitations and biases of different measurements \citep{Ettori2013b, Pratt2019}. 

In choosing our method, we prioritized obtaining reliable and converged mass estimates for over $10^4$ haloes, spanning more than 3 orders of magnitude in mass. 
For each halo, we bin all gas cells $>10^5$\,K in 25 linearly-spaced radial bins extending from the halo center all the way to $1.5$\,\rcrit\,. 
We compute the gas density profile from the median surface brightness estimated for each radial bin, and we sum the pixel spectra in each bin. 
The resulting X-ray spectrum is modelled as a single-temperature plasma with three free parameters: temperature, emission measure and metallicity.
For each radial bin, 
we perform the fit in the  0.5 -- 10\,keV energy range, utilizing spectra interpolated from the \textsc{apec} \citep{Smith2001} lookup table. 
We thus obtain a radial profile of the spectroscopic temperature, $T_{\rm X}(r)$, as a function of the cluster-centric radius $r$.

The total mass enclosed in a sphere of radius $r$ can be derived from the following solution of the hydrostatic equilibrium equation:
\begin{equation}
M_{\rm tot} (<r) = - \frac{r\, k_B \, T(r)}{\mu\, m_p \,G} \left( \frac{{\rm d \, ln}\, \rho_g(r)}{{\rm d \, ln}\, r} + \frac{{\rm d \, ln}\, T(r)}{{\rm d \, ln}\, r}\right),\label{eqn:hydroeq}
\end{equation}
where $k_{\rm B}$ is the Boltzmann constant, $G$ is the gravitational constant, $\mu$ is the mean molecular weight, $m_p$ is the proton mass, $r$ is the radius from the center of the halo, and $T(r)$ is the mean temperature at radius $r$.
In order to compute the gas density ($\rho_g$) and temperature ($T$) gradients, we turn to the method described in \citet{Vikhlinin2006}. 
Thus, we assume the three-dimensional gas density distribution can be described by a modified $\beta$-model \citep{Cavaliere1978}. This modification includes modelling the centers of clusters using a power law for the cusp \citep[e.g.,][]{Pointecouteau2004} and adjusting the slope by a term $\varepsilon$ and transition region $\gamma$ at large radii ($r_s \gtrsim 0.3\, R_{\rm 200crit}$ from \citet[][]{Vikhlinin1999, Neumann2005}) relative to the best-fit slope used to model the profile at small radii. The model thus takes the form:
\begin{equation}\label{eqn:eqndensitymodel}
n_e n_H = n_0^2 \frac{(r/r_c)^{-\alpha}}{(1+r^2/r_c^2)^{(3\beta - \alpha /2)}} \frac{1}{(1+r^\gamma/r_s^\gamma)^{\varepsilon/\gamma}}.\end{equation}
Compared to equation~(3) in \citet{Vikhlinin2006}, we ignore the secondary $\beta$-model component with a smaller core radius, which was proposed as a way of adding additional flexibility in modelling the cluster centers. The $\gamma$ parameter is fixed to a value of $\gamma = 3$, while the other parameters are free. Additionally, $\varepsilon$ is constrained to values of $\varepsilon < 5$, which avoids density breaks that are unphysically sharp \citep{Vikhlinin2006}. 

The three-dimensional temperature profile is modelled following equation (6) in \citet{Vikhlinin2006}:
\begin{equation}\label{eqn:eqntempmodel}
    T_{\rm 3D} (r) = T_0\, t_{\rm cool}(r) \,t(r)\,,
\end{equation}
where the temperature profile is composed of a broken power law term:
\begin{equation}
    t(r) = \frac{(r/r_t)^{-a}}{\left[ 1 + (r/r_t)^b\right]^{c/b}},
\end{equation}
multiplied by an additional term:
\begin{equation}
t_{\rm cool} = \frac{(x + T_{\rm min} / T_0)}{(x + 1)} \;\;\;\; {\rm where} \;\;\;\; x = \left(\frac{r}{r_{\rm cool}}\right)^{a_{\rm cool}},
\end{equation}
that describes the temperature decline that is observed at the centers of most clusters due to radiative cooling, as described in \citet{Allen2002}.
Both the density and temperature models described by equations~(\ref{eqn:eqndensitymodel}) and (\ref{eqn:eqntempmodel}) have great functional freedom and we find that they accurately model the relatively smooth density and temperature profiles of the $>10^4$ haloes in our sample. 

\section{Hydrostatic and X-ray Mass Bias}
\label{sec:massbias_yxysz}

\begin{figure*}
\centering
\includegraphics[width=0.99\textwidth]{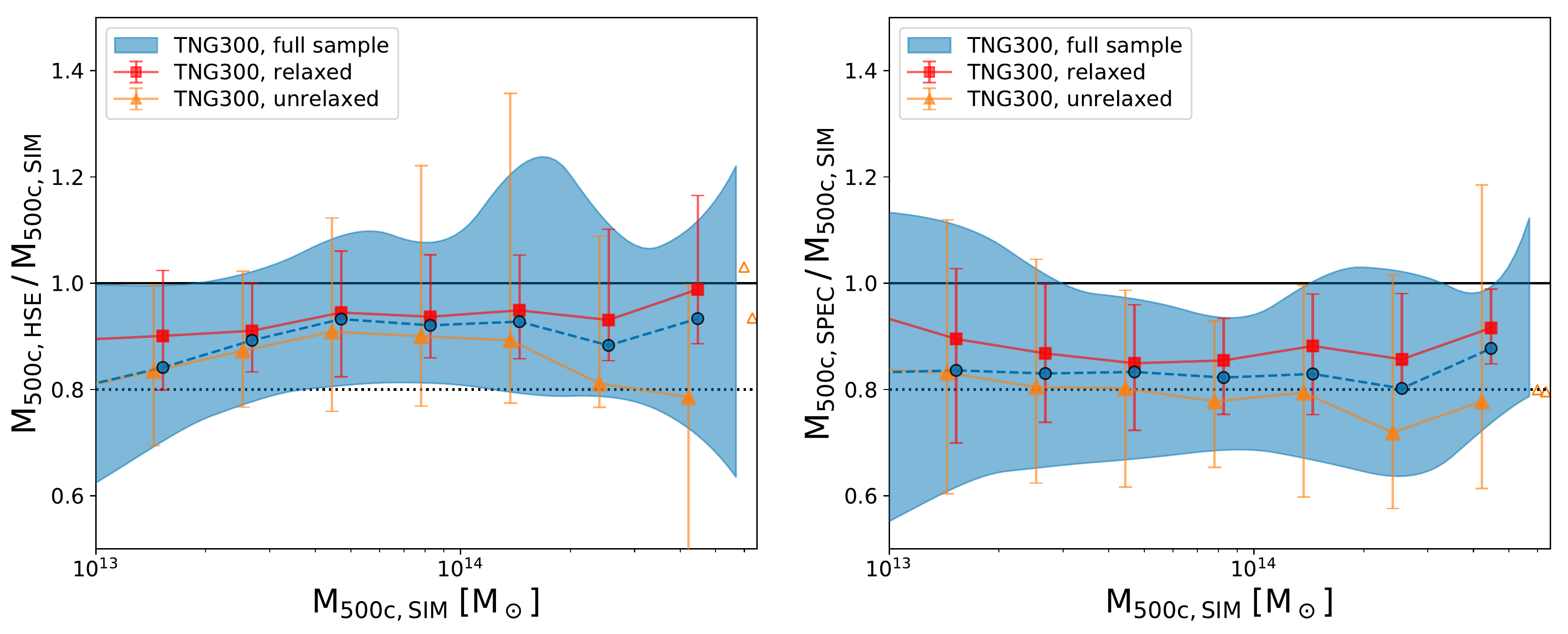}
\caption{Ratio between the hydrostatic mass estimate and the true mass (\msim) as a function of halo mass, for the $z=0$ sample from TNG300.
We show median values for the full sample (blue circles), relaxed sample (red squares) and unrelaxed sample (orange triangles) for all mass bins with at least five objects. Values for the remaining high-mass unrelaxed clusters are indicated with open orange triangles. 
The \textit{left panel} showcases hydrostatic mass estimates ($M_{\rm 500c, \, HSE}$) derived from true simulation density and temperature profiles. In the \textit{right panel}, we present estimates for X-ray masses derived from synthetic X-ray images for the same haloes as those in the left panel. 
Blue contours mark the 16\textsuperscript{th} and 84\textsuperscript{th} percentiles for the full sample from TNG300, with the continuous edges calculated using quadratic interpolation. Error bars in both panels correspond to 16\textsuperscript{th} and 84\textsuperscript{th} percentiles for the relaxed (red) and unrelaxed samples (orange), respectively. Hydrostatic mass estimates have an average mass bias for haloes with \msim\, $\geq 10^{13}$ \ms\, of ${\rm b}_{\rm HSE} = 1 -$ \mhse\,/\msim\, $= 0.125^{+0.004}_{-0.003}$ for the full sample. Nonetheless, there is significant scatter around the average mass bias, with a non-negligible fraction of high mass haloes exhibiting $M_{\rm 500c, \, HSE}$ > \msim. In contrast,  
X-ray mass estimates show a significantly higher degree of bias, with an average mass bias  of ${\rm b}_{\rm X-RAY} = 1 -$ \mspec\,/\msim\, $= 0.170\pm0.004$ for the full sample above \msim \,$\geq 10^{13}$ \ms. On average, relaxed haloes have smaller mass biases than unrelaxed haloes, irrespective of the choice of mass estimate (\mhse\, or \mspec).}
\label{fig:Mbiashydrospec}
\end{figure*}

One of the major sources of uncertainty in cluster cosmology comes from estimating halo masses. For the rest of this paper, we define the mass bias as:
\begin{equation}
b = 1 - \frac{M_{\rm EST}}{M_{\rm TRUE}},
\end{equation}
where we define $M_{\rm TRUE} \equiv \,$\msim\, to be the total mass enclosed in a sphere of radius \rsim\, (as identified by \textsc{subfind}),  centered on the gravitational potential center of the halo. 
To estimate masses inside the spectroscopic aperture, \rspec\,, we utilize the thermodynamic profiles derived from the \textsc{Mock-X} pipeline (Section \ref{subsec:xraypipeline_yszmyx}) together with the hydrostatic equilibrium equation (\ref{eqn:hydroeq}). The spectroscopic aperture \rspec\, is defined as the intersection point between the X-ray-derived total density profile and $500 \times \rho_{\mathrm{crit}}$. Lastly, \mspec\, is the total halo mass contained within the spectroscopic aperture. 
In our current study, we explore different estimated masses,  $M_{\rm EST}$, computed under the assumption of hydrostatic equilibrium (Section~\ref{sec:massbias_yxysz}), as well as by utilizing X-ray observables as mass proxies (Section~\ref{sec:yxmassproxy}).

\subsection{Hydrostatic Mass Bias}
\label{subsec:hydromassbias}

In Figure \ref{fig:Mbiashydrospec}, we present how mass biases in \tng\, evolve as a function of total halo mass.
For a mass cut of \mcrit\, $=10^{14}$\,\ms\,, we find an average bias  $b=0.099^{+0.008}_{-0.026}$ for the full sample. 
For relaxed clusters, the average bias is reduced to only $b = 0.052^{+0.016}_{-0.022}$, while the bias increases for unrelaxed objects: $b=0.122^{+0.023}_{-0.015}$. The errors for the median bias values were computed by generating 10,000 bootstrap resamples. 
Overall, we find a median hydrostatic mass bias of $\sim 10\%$ for the full sample and $\sim 5\%$ for relaxed clusters, but we observe a large level of scatter around this median value for $b$. 
In Figure \ref{fig:Mbiashydrospec}, we show contours marking $1\sigma$ scatter around the median bias for the full sample, as well as errorbars marking the 16\textsuperscript{th} and 84\textsuperscript{th} percentiles around the median bias for relaxed haloes (red) and unrelaxed haloes (orange), respectively. 
We find a very large level of scatter in the hydrostatic bias, with $1\sigma$ contours consistently extending beyond the $b \in [0, 0.2]$ range. In particular, for the highest mass haloes in our sample (\mcrit \,$\geq 10^{14}$\,\ms), we find several clusters that have negative hydrostatic mass bias, i.e., \mhsec\, > \msimc. 

Based on the median bias values, it appears that the hydrostatic mass bias decreases with halo mass, especially for relaxed clusters.
The average bias for the full sample drops from $b = 0.125^{+0.004}_{-0.003}$ for a mass cut of $10^{13}$\,\ms\, down to $b = 0.099^{+0.008}_{-0.026}$ for \mcrit\, $>10^{14}$\,\ms. For the relaxed sample, the bias similarly drops by about $0.02$, from $b = 0.070^{+005}_{-0.004}$ for \mcrit\, $>10^{13}$\,\ms\, down to  $b = 0.052^{+0.016}_{-0.022}$ above $10^{14}$\ms. 

For the unrelaxed sample, we see a reduction in bias within the mass range \linebreak \mcrit\,$\in[5\times10^{13}$\,\ms\,--\,$10^{14}$\,\ms], coupled with an increase in bias for the most massive clusters in our sample. This increase in the hydrostatic mass bias of massive unrelaxed clusters is also accompanied by a significant increase in scatter, as can be observed from the increasingly larger error bars in Figure~\ref{fig:Mbiashydrospec}. As a result of recent major mergers, some of these more massive clusters are likely to deviate strongly from hydrostatic equilibrium, which would explain the increase in both median bias and its scatter for the highest mass haloes.
One likely origin of the hydrostatic mass bias is the non-thermal pressure provided by bulk and turbulent gas motions in clusters \citep{Lau2009, Lau2013, Nelson2012, Nelson2014a, Nelson2014b, Shi2015, Shi2016}. The non-thermal pressure component is also expected to scale smoothly with haloes mass down to galaxy groups \citep{Green2020}. 

\subsection{X-ray Mass Bias}
\label{subsec:xraymassbias}
In the right panel of Figure~\ref{fig:Mbiashydrospec}, we show the equivalent mass estimates derived using gas density and temperature profiles computed from the mock X-ray analysis. 
We find a significant increase in the measured bias, with a median bias of $b = 0.170^{+0.004}_{-0.004}$ for the full sample above \mcrit $= 10^{13}$\,\ms. This is $0.045$ higher than the bias computed for the same sample using the true density and temperature profiles  ($b_{\rm HSE} = 0.125^{+0.004}_{-0.003}$). 
Once again, we find that relaxed haloes tend to have smaller mass biases, with $b_{\rm X-RAY}^{\rm relaxed} = 0.139^{+0.004}_{-0.006}$ at \mcrit\, $>10^{13}$\,\ms. However, $b_{\rm X-RAY}^{\rm relaxed}$ is twice the hydrostatic mass bias for relaxed clusters:  $b_{\rm HSE}^{\rm relaxed} = 0.070^{+0.005}_{-0.004}$. 
X-ray mass estimates exhibit similar scatter as the hydrostatic mass estimates, but the scatter in X-ray masses increases at smaller halo masses.
This may be related to the greater degree of uncertainity in measuring spectroscopic temperatures for low-mass galaxy groups. For \msim\, $\simeq 10^{13}$\,\ms\,, $1\sigma$ contours for the ratio of X-ray mass estimates to true masses range from \mspecc\,/\msimc\, $ \simeq 1.1$ all the way down to \mspecc\,/\msimc\, $ \simeq 0.5$. 

The median values of the X-ray mass bias do not show a significant trend with halo mass. The exception comes from unrelaxed clusters, for which the X-ray mass bias increases slightly with mass, just as it did in the case of the hydrostatic mass bias (see Section~\ref{subsec:hydromassbias}). Overall, the level of scatter in $b$ from halo to halo dominates over any mass dependence for the median bias value.

Previous studies of cosmological simulations highlighted significant differences in the X-ray mass bias estimated with smooth-particle-hydrodynamics (SPH) and adaptive-mesh-refinement (AMR) codes \citep[e.g.,][]{Rasia2014}. 
The mass bias found in IllustrisTNG is in reasonably good agreement with the \textsc{BAHAMAS} and \textsc{MACSIS} simulations. For the \textsc{BAHAMAS} sample, the hydrostatic mass bias is  $b=0.13 \pm 0.002$ on a sample of clusters $> 10^{14}$\,\ms\, and the \textsc{MACSIS} sample yields $b=0.15\pm0.003$ but it only includes clusters above $4\times10^{14}$\,\ms\, \citep{Barnes2021}. Clusters in TNG300 (\msim\,$> 10^{14}$\,\ms) have slightly lower hydrostatic mass bias, $b = 0.099^{+0.008}_{-0.026}$, than either the \textsc{BAHAMAS} or \textsc{MACSIS} samples, but this is likely explained by the higher number of haloes with \msim\,$\approx10^{14}$\,\ms\, in the IllustrisTNG sample compared to the larger fraction of high-mass clusters in the samples of the other two simulations. 
\citet{Barnes2021} report that the \textsc{BAHAMAS} sample yields a significantly lower X-ray mass bias, $b = 0.11 \pm 0.003$, than the \textsc{MACSIS} sample, $b= 0.25\pm0.005$. 
The X-ray mass bias in TNG300 lies in-between the values predicted by \textsc{BAHAMAS} and \textsc{MACSIS}, with a median value  $b= 0.174^{+0.022}_{-0.010}$ for \msim\,$> 10^{14}$\,\ms. Despite the general agreement between simulations, in the era of precision cosmology, we need to better understand the source of the remaining discrepancies in the  X-ray mass bias predicted by different physical models.  

\section{Mass Dependence of \yx\, and \ysz\,}
\label{sec:yxyszmassevolution}

\begin{figure*}
\centering
\includegraphics[width=0.99\textwidth]{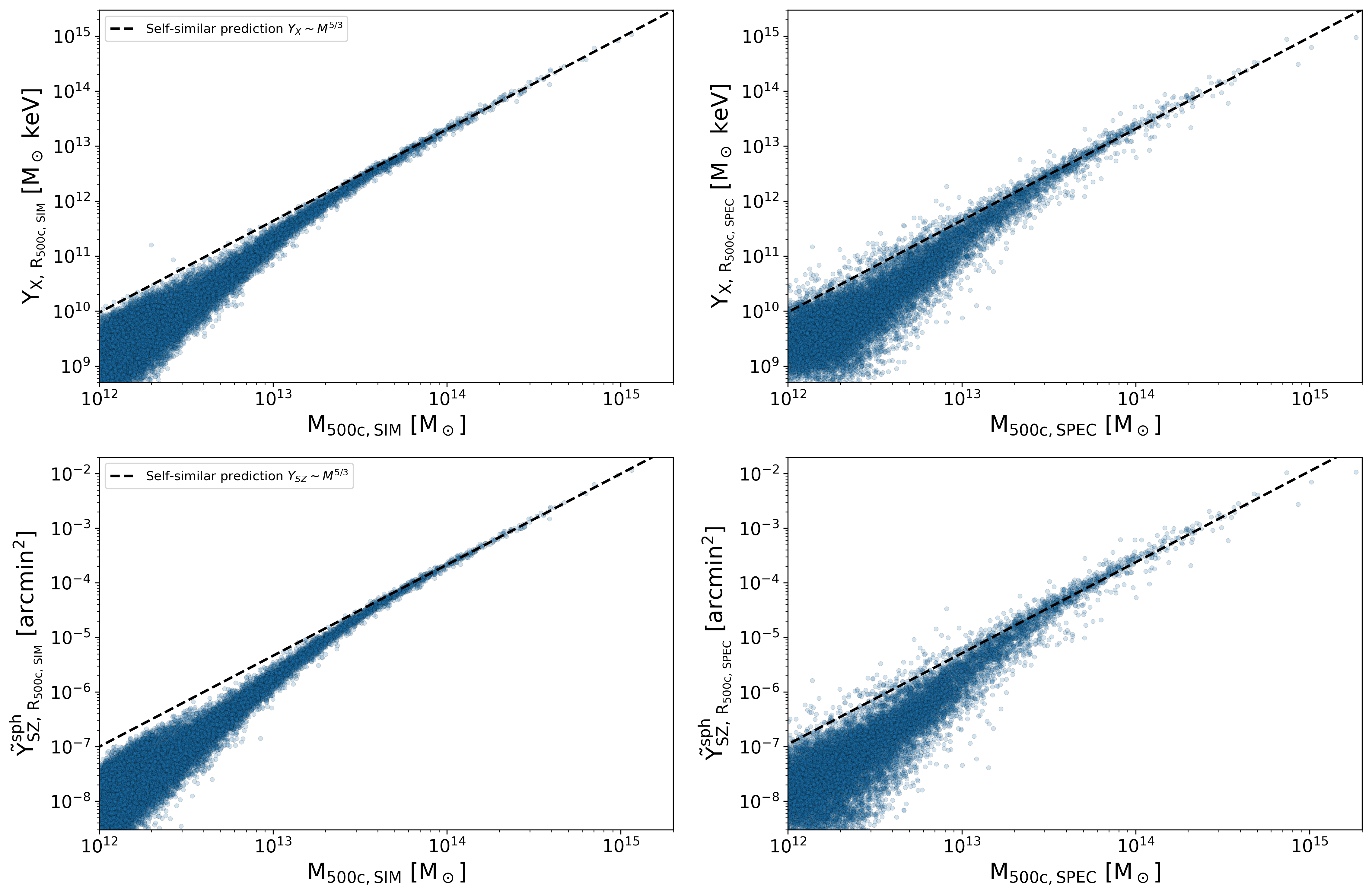}
\caption{Mass dependence of the \yx\, (top panels) and \ysz\, (bottom panels) signal for all objects with \mcrit\, $\geq 10^{12}$ \ms\, in the $z=0$ TNG300 simulation of the IllustrisTNG suite. 
We compare the level of scatter seen in the left column for quantities measured inside the true simulation aperture (\rsim), which gives relatively tight \yxsim\, -- \msim\, and \yszsim\, -- \msim\, relations, to the significantly larger scatter in the right column for quantities measured inside the spectroscopic aperture (\rsim). 
Black dashed lines correspond to the self-similar expectation for the scaling relations, i.e., \yx\, $\propto M^{5/3}$ and \ysz\, $\propto M^{5/3}$ at all mass scales. The normalizations of the self-similar lines were determined by computing the best-fit simple power law with slope fixed to $\alpha = 5/3$, evaluated on the full sample of high-mass clusters \mcrit\, $\geq 10^{14}$ \ms\, at $z=0$. We note that the self-similar line forms an upper envelope encompassing almost all small scale haloes for quantities measured inside \rsim. Due to the higher degree of scatter we find for quantities measured inside \rspec, a few more galaxy groups scatter above the self-similar line, but we still find that fewer than than 1\% of haloes below \mbox{\mcrit \, $\geq 10^{13}$ \ms\,} have \yx\, and \ysz\, signals above the self-similar scaling calibrated on the largest clusters. There is a clear indication for a break from self-similarity in both the \yx\, -- $M$\, and \ysz\, -- $M$ scaling relations in IllustrisTNG simulation.}
\label{fig:scatterYxYsz}
\end{figure*}

In Paper I, we found that \yx\,-- $M$ and \ysz\,-- $M$ are two of the tightest scaling relations in IllustrisTNG across three orders of magnitude in \mcrit.

Figure~\ref{fig:scatterYxYsz} presents all $z=0$ haloes with \mbox{\mcrit\, $\geq 10^{12}$\,\ms} from the TNG300 simulation.
The black dashed lines mark the self-similar expectation for each of the scaling relations, i.e., \yx\, $\propto M^{5/3}$ and \mbox{\ysz\, $\propto M^{5/3}$} at all mass scales. 
The normalization for each self-similar line was determined by computing the best-fit simple power law with slope fixed to $\alpha = 5/3$, evaluated on the individual data points corresponding to the highest-mass $z=0$ clusters (\mcrit\, $\geq 10^{14}$\,\ms) in our sample.
The spread of the individual data points emphasizes the large increase in intrinsic scatter when we switch from quantities (such as $M$, \yx, \ysz) measured inside the true simulation aperture \rsim\, (left panel) to quantities measured inside the spectroscopic aperture \rspec\, (right panel). This large halo-to-halo variation in the ratio of \mspec\,/\msim\, was also apparent from our discussion on the X-ray mass bias in Figure~\ref{fig:Mbiashydrospec}.

The self-similar lines form upper envelopes encompassing almost all galaxy groups (\mcrit\,$\lesssim 4\times 10^{13}$\,\ms) in \tng, for quantities measured inside \rsim.  Due to the higher degree of scatter in quantities measured inside \rspec, a few more galaxy groups scatter above the self-similar line, but we still find that fewer than 1\% of haloes below \mbox{\mcrit \, $\leq 10^{13}$ \ms\,} have \yx\, and \ysz\, signals above the self-similar simple power law model calibrated on the largest clusters. 
 
Data from IllustrisTNG suggest a break from self-similarity for both the \yx\, -- $M$\, and \ysz\, -- $M$ scaling relations.
To model these scaling relations, we use the simple power law (SPL) and smoothly broken power law (SBPL) models introduced in Paper I. 

The SPL model is given by:
\begin{equation}
    Y = 10^A \left( \frac{X}{X_{\rm norm}}\right)^\alpha \label{eqn:eqnSPL}
\end{equation}
with two free parameters ($A$, $\alpha$) for the normalization and slope, respectively. 

The SBPL model, on the other hand, is defined as:
\begin{equation}
\frac{Y(X)}{Y(X_{\rm norm})} =  \left( \frac{X}{X_{\rm norm}} \right)^{ \left(\frac{\alpha_2 + \alpha_1}{2}\right)} \left[ \frac{{\rm cosh} \left(\frac{1}{\delta} {\rm log}_{10} \left( \frac{X}{X_P}\right) \right)  }{ {\rm cosh} \left( \frac{1}{\delta} {\rm log}_{10} \left(\frac{X_{\rm norm}}{X_p}\right)\right) } \right]^{\left( \frac{\alpha_2 - \alpha_1}{2}\right) \, \delta \, {\rm ln} 10},\label{eqn:eqnSBPLletter}
\end{equation}
where the exponent $\alpha_{\rm SBPL}$ is a function of $X$ defined as:
\begin{equation}
\alpha_{\rm SBPL} = \frac{{\rm d} \log Y}{{\rm d} \log X} = \frac{\alpha_2 - \alpha_1}{2} \tanh \left[ \frac{1}{\delta} {\rm log}_{10} \left(\frac{X}{X_p} \right) \right] + \frac{\alpha_2 + \alpha_1}{2}.\label{eqn:eqnSBPLslope_letter}
\end{equation}
Here, $\delta$ is a measure of the width of the transition region between the two slopes ($\alpha_1$ at low masses and $\alpha_2$ at high masses, respectively). In the limit of $\delta \rightarrow 0$, the SBPL model is identical to a broken power law model with free pivot \xp.  Summaries of the best-fit scaling relations based on the SPL and SBPL models are included in Table~6  and Table~7 in the Appendix of Paper I.

Note that the local slope ($\alpha_{\rm SBPL} = {\rm d}\,Y\,/\,{\rm d}\,X$ in eqn.~\ref{eqn:eqnSBPLslope_letter}) for the SBPL model is relatively simple, only including a hyperbolic tangent term that allows the break to occur smoothly over a range of halo masses. However, the full expression for $Y_{\rm SBPL} (X)$, as defined in equation~(\ref{eqn:eqnSBPLletter}), is more complex, since the unknown $X$ (or halo mass) value appears both inside the hyperbolic cosine terms, as well as in the first term ($X/ X_{\rm norm}$). We therefore do not use an analytic expression for the inverse of this equation, but instead we compute the numerical inverse of equation~(\ref{eqn:eqnSBPLletter}) for the best-fit SBPL model evaluated over the entire mass range of haloes included in our sample. Moreover, rather than computing the numerical inverse of the SBPL, an approximate solution would be to use the analytic inverse of the equation for a BPL with a free pivot:
\begin{equation}
    X =
    \begin{cases}
    X_{\rm norm} \left( \frac{Y}{10^{A_1}} \right)^{\frac{1}{\alpha_1}} \text{, } X < \,  X_{\rm pivot}\\
    X_{\rm norm} \left( \frac{Y}{10^{A_2}} \right)^{\frac{1}{\alpha_2}} \text{, } X \geq \,  X_{\rm pivot}
    \end{cases}
\end{equation}
where \xp\ corresponds to a pivot location for the observable quantity (in this case, \yx) which is by construction equal to:
\begin{equation}
Y_{\rm pivot} = 10^{A_1} \left( \frac{X_{\rm pivot}}{X_{\rm norm}} \right)^{\alpha_1} = 10^{A_2} \left( \frac{X_{\rm pivot}}{X_{\rm norm}} \right)^{\alpha_2}.
\end{equation}

We note that the parameter characterizing the transition width, $\delta$, is relatively small for the best-fitting SBPL for the \yx\,--\mcrit\, scaling (see Table~6 in the Appendix of Paper 1). For the full sample, we find a best-fitting value of $\delta = 0.13^{+0.03}_{-0.04}$ for \rsim\, aperture and $\delta=0.14^{+0.09}_{-0.07}$ for \rspec\, aperture. The width parameter becomes significantly larger if we restrict the analysis to relaxed clusters, where we find $\delta = 0.49^{+0.73}_{-0.29}$ and $\delta = 1.46^{+1.10}_{-0.88}$ for the simulation and spectroscopic apertures, respectively. Thus, the analytic inverse to the BPL model can provide a reasonable approximation for a sample dominated by unrelaxed objects, but we caution against using it for samples that include relaxed clusters.

We find a remarkable level of agreement between the location of the break in the \yxm\, and \yszm\, scaling relations, with \xp = $3.8 \pm0.2 \times 10^{13}$\,\ms\, for \yx\, and \xp = $3.7 \pm0.1 \times 10^{13}$\,\ms\, for \ysz. While the \yx\, parameter was introduced by \citet{Kravtsov2006} with the intent of creating an X-ray equivalent of the \ysz\, parameter, it was not necessarily expected that the breaks in the two scalings would coincide. \ysz\, is proportional to the mass-weighted temperature, while \mbox{\yx\, $\equiv T_{\rm X,\, ce} \times M_{\rm g}$} is proportional to the spectroscopic temperature. The agreement between the break location in both scalings suggests that the two temperatures ($T_{\rm mw}$ and $T_{\rm X,\, ce}$) have a similar dependence on halo mass. 

\ysz\, and \yx\, disagree slightly in the best-fit slopes at low masses. We can also notice this effect by inspecting the left panels of Figure~\ref{fig:scatterYxYsz}, where it is apparent that the \ysz\, signal deviates from self-similarity more strongly at the lowest mass scales ($\alpha_1 = 2.418^{+0.004}_{-0.003}$) than the slope predicted by \yx\, ($\alpha_1 = 2.260^{+0.003}_{-0.003}$). On the other hand, \yxm\, and \yszm\, once again show good agreement in their predictions for the highest mass clusters in \tng. The slope predicted at the high mass end by \yx\, ($\alpha_2 = 1.710^{+0.021}_{-0.018}$) is well within  $1\sigma$ from the slope predicted by \ysz\, ($\alpha_2 = 1.687^{+0.015}_{-0.015}$). The uncertainty for each of the parameters was estimated through bootstrapping with replacement using 10,000 bootstrap resamples.  
When switching to a spectroscopic aperture, \rspec\,, the additional scatter is also accompanied by small biases in the best-fit parameters. Our results indicate that the break shifts to higher masses for both scaling relations, and the effect is more pronounced for \yx\, (\xp $= 7.2 \times 10^{13}$\ms) than for \ysz\, (\xp $= 5.2 \times 10^{13}$\ms). The slight discrepancies we find between the \yszm\, and \yxm\, scaling relations may arise from biases in X-ray derived measurements of $M_{\rm gas}$ and $T_{\rm X}$ compared to the true gas density and temperature profiles of individual haloes. 

Interestingly, the good agreement in the pivot location does not hold true for the relaxed sample. Relaxed clusters prefer a high-mass break in \yxm\, (\xp $= 1.1 \times 10^{14}$\,\ms) compared to the break for \yszm\, (\xp $= 5.5 \times 10^{13}$\,\ms). In both cases, the relaxed sample leads to a break point in the scaling that is biased high compared to the full sample. 
In particular, the most massive relaxed clusters in TNG300 follow a power law consistent with the self-similar slope for the \yszm\, relation and this result holds true irrespective of the chosen aperture (for \yszsim\,-- \msim\,, $\alpha_2 = 1.696^{+0.026}_{-0.052}$, and for \yszspec\,-- \mspec\,, $\alpha_2 = 1.667^{+0.03}_{-0.04}$). 
On the other hand, \yx\, for relaxed clusters in TNG leads to a somewhat shallower best-fit slope at high masses: $\alpha_2 = 1.581^{+0.160}_{-0.581}$ for \yxsim--\msim\, and $\alpha_2 = 1.425^{+0.237}_{-0.109}$ for \yxspec\,-- \mspec. 
Moreover, the uncertainty around the best-fit slopes for \yxm\, is much higher for the relaxed scaling than for the unrelaxed sample, and the self-similar slope of $1.67$ is about $1 \sigma$ above the best-fit values for $\alpha_2$.

To summarize, our results confirm that \yx\, and \ysz\, share many similarities, including the slopes they predict for scaling relations of very massive clusters, and the location of the break in the scalings. 
However, the points where we find slight discrepancies between the two scaling relations offer promising avenues to further investigate the physical processes responsible for these differences. 
For galaxy groups and galaxies, the slopes predicted by \yx\, and \ysz\, begin to deviate, with \ysz\, decreasing quicker (steeper slope) with decreasing mass than \yx. Our findings also suggest that \yx\, is more sensitive to differences in the dynamical state of the cluster compared to \ysz. While relaxed clusters are still following a self-similar \yszm\, scaling, their slopes in \yxm\, are biased to lower values (i.e., shallower slopes).

\section{Robust Mass Proxies for SZ and X-ray surveys} 
\label{sec:unifying}

In order to take advantage of the full power of \ysz\, measurements to constrain cosmological parameters, observers need accurate calibrations for the total masses of haloes in their samples. Ideally, the methods being used should minimize systematic errors and uncertainties in connecting the  \ysz\, signal to the total halo mass, \mtot.  In order to connect the \ysz\, signal to the total halo mass, \mtot\,, observers need to calibrate the masses of haloes in their samples using methods that reduce any systematic errors or uncertainties in \mtot. 
The tight correlation in the \yxm\, scaling relation has led \yx\, to be proclaimed as the most robust X-ray mass estimator \citep[][]{Kravtsov2006, Vikhlinin2009a}. Nonetheless, most studies that aim to put constraints on cosmological parameters still assume a simple power law \yxm\, scaling relation. For example, the \citet{PlanckCollaboration2014b} calibrates the \yx\,--mass proxy against hydrostatic masses \citep{Arnaud2010}, assuming a self-similar mass dependence in the scaling relation used for calibration \citep[][]{PlanckCollaboration2014a, PlanckCollaboration2014b}.
In this section, we present a novel approach for estimating cluster masses for SZ surveys, based on the robust mass proxy \myx. 

\subsection{\yx\, Mass Proxy}
\label{sec:yxmassproxy}

First, we discuss several common approximations for the dependence of \yx\, on halo mass. We start with the simplest (and most naive) assumption that \yx\, scales self-similarly with mass:
\begin{equation}
   \textit{Assumption 1:} \;\;\;\; Y_{\rm X} \propto M^{5/3}
\end{equation}
for all halo masses. This scaling can be inverted in order to obtain the functional form of the \yx\,--mass proxy, denoted by $M_{{\rm Y}_{\rm X}}$. Under this assumption, \myx\, is simply evolving with mass according to \myx\, $\propto Y_{\rm X}^{3/5}$.

Physical processes such as AGN feedback launch powerful jets in the ICM that can heat the surrounding gas, increasing its temperature. Deviations from the self-similar slope in \yxm\, could be possible depending, for example, on the exact power and duty cycle of AGN feedback. Thus, we next explore the scenario in which \yxm\, still follows a simple power law, but the slope is biased with respect to the self-similar prediction:
\begin{equation}
    \textit{Assumption 2:} \;\;\;\; Y_{\rm X} \propto M^{\alpha_{\rm SPL}}
\end{equation}
For a mass cut of $10^{12}$\,\ms\,, we perform a simple power law on the geometric means of the \yxm\, relation and we find a best-fit slope of $\alpha_{\rm SPL} = 2.01$. Under \textit{Assumption 2}, \myx\, should thus evolve with mass according to \myx\, $\propto Y_{\rm X}^{0.498}$.

The final assumption we will explore involves characterising the \yxm\, scaling relation with a model that allows more flexibility in the observed trend between \yx\, and halo mass.  Specifically, we propose a SBPL model to capture the break in the \yxm\, scaling relation which occurs at group scales, as well as the transition range around this break. Thus, the third assumption for modeling the \yx--mass proxy is derived from fitting a SBPL to the \yxm\, scaling relation:
\begin{equation}
    \textit{Assumption 3:} \;\; Y_{\rm X} \propto Y_{\rm SBPL} (M),
\end{equation}
with  $Y_{\rm SBPL} (M)$ following the model defined in equation~(\ref{eqn:eqnSBPLletter}).

\subsection{\ysz\,-- \myx\, Scaling Relation}
\label{sec:yszmyx}

Next, we explore the bias introduced by the \yx\,--mass proxy on the derived \ysz\,--\,$M_{Y_{\rm X}}$ scaling, for a variety of underlying models for the \yx\,--mass proxy function, $M_{Y_{\rm X}} (M)$, with the goal of quantifying how much different assumptions for the \yxm\, scaling relation can bias the inferred \ysz\, -- \myx\, data away from the true \yszm\, scaling relation. 

\begin{figure*}
\centering
\includegraphics[width=0.99\textwidth]{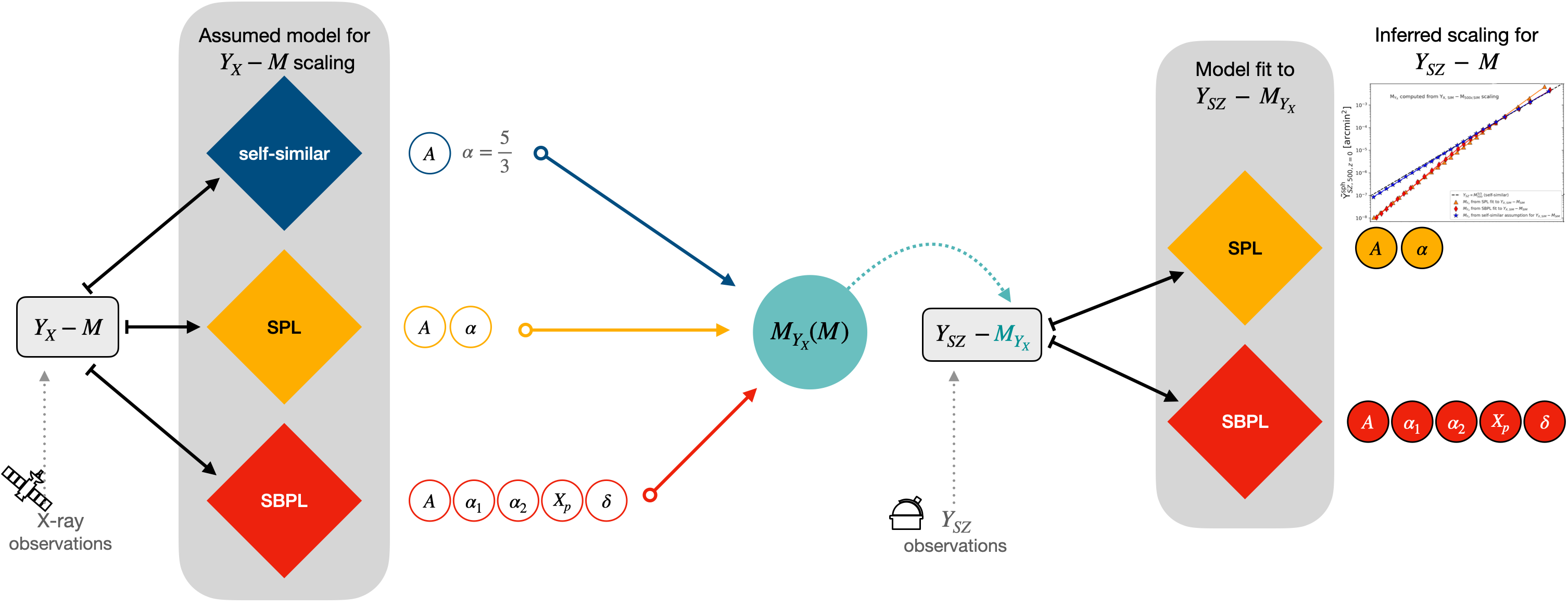}
\caption{Diagram describing the modelling for the \ysz\,--\,\myx\, scaling, based on observational input data for \yx\, and \ysz. The \yx\,--\,mass proxy (\myx) can be derived by assuming an underlying model for the \yxm\, scaling relation. Some common choices include a self-similar model (\yx\, $\propto 10^A \, M^{5/3}$) or a simple power law (\yx\, $\propto 10^A \, M^{\alpha}$). 
We show that the SBPL model provides a significantly better fit for the break in the scaling relation. This model includes a normalization parameter $A$, the best-fit slopes at small mass scales ($\alpha_1$) and high mass scales ($\alpha_2$), as well as a pivot for the break (\xp) and a parameter that characterizes the width of the transition region around the break ($\delta$). Based on the assumed model (e.g., self-similar, SPL, SBPL), one can derive a functional form for the \yx\, mass proxy as a function of halo mass: \myx(M). Finally, one can combine the observed X-ray and SZ data to derive the scaling relation between the observed total SZ signal, \ysz\,, and the \yx\,--\,mass proxy, \myx. The final inferred best-fit parameters will be highly dependent on the assumed model for the \yxm\, scaling relation. As shown in Figure \ref{fig:Ysz_Myx_bestfits}, assuming that \yxm\, follows the self-similar prediction will in turn almost entirely mask the true break in the \yszm\, scaling. In addition, a simple power law fit for  the \yxm\, scaling also biases the resulting  \ysz\,--\,\myx\, relation, generating an artificially steep slope. A SBPL for \yxm\, will recover, with good precision, the location of the break in the underlying \yszm\, scaling relation.}
\label{fig:diagramYszletter}
\end{figure*}

In Figure~\ref{fig:diagramYszletter}, we present a schematic diagram of the approach we adopt in this study. 
The \yx\,--mass proxy (\myx) can be derived by assuming an underlying model for the \yxm\, scaling relation. Some common choices include a self-similar model (\yx\, $\propto 10^A \, {\rm M}^{5/3}$) or a simple power law (\yx\, $\propto 10^A \, {\rm M}^{\alpha}$). 
In this work, we show that a SBPL model provides a significantly better fit for the break in the scaling relation of \yxm. This model includes a normalization parameter $A$, the best-fit slopes at small mass scales ($\alpha_1$) and high mass scales ($\alpha_2$), as well as a pivot for the break (\xp) and a parameter that characterizes the width of the transition region around the break ($\delta$). Based on the assumed model (e.g., self-similar, SPL, SBPL), one can derive a functional form for the \yx\, mass proxy as a function of halo mass, as we described in the previous section.
Finally, one can combine the observed X-ray and SZ data (or in our case, the values of \yx\, and \ysz\, estimated for our sample of simulated haloes) to derive the scaling relation between the observed total SZ signal, \ysz\,, and the \yx\,--mass proxy, \myx. The final inferred best-fit parameters will be highly dependent on the initial assumption that was made in order to relate \yx\, to the total halo mass within \rcrit.

\begin{figure*}
\centering
\includegraphics[width=0.99\textwidth]{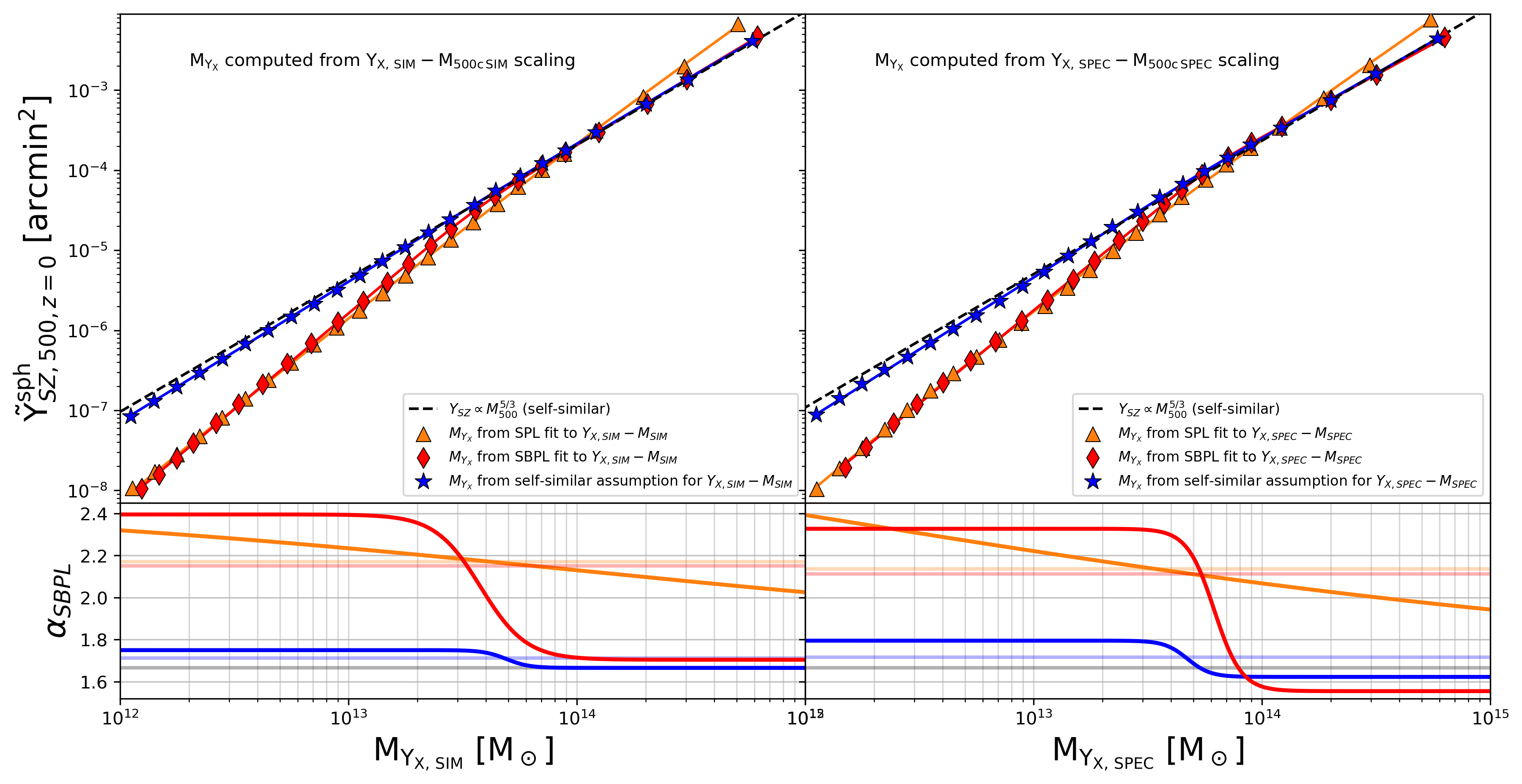}
\caption{Best-fits for the \ysz\,--\,\myx\, scaling relations. The \yx--\,mass proxy (\myx) was derived from the \yxm\, relation, with \yx\, and $M$ measured inside a true simulation aperture \rsim\, (left column) and inside a spectroscopic aperture \rspec\, (right column). 
The underlying assumptions for the \yxm\, scaling relation will lead to biases in the recovered slopes for the \ysz\,--\,\myx\, scaling. Making the simplifying assumption that \yx\, scales with mass as a simple power law (orange triangles) biases high the \ysz\, signal at cluster scales (predicted slope is steeper than true \yszm\, slope). 
Blue stars mark geometric means for the \ysz\,--\,\myx\, relation when \myx\, is derived by assuming \yx\, scales self-similarly with mass, i.e., \yx\, $\propto M^{5/3}$ at all mass scales.
The dashed black line represents the self-similar scaling between the SZ signal and the true simulation mass: \ysz\, $\propto M^{5/3}$ at all mass scales, with the normalization fixed to the best-fit value obtained from a simple power law fit on the highest mass clusters \mcrit\, $\geq 10^{14}$\,\ms\, at $z=0$. We note that assuming self-similarity for \yxm\, leads to an almost self-similar result for the \ysz\,--\,\myx\, scaling. 
Nonetheless, the true break in the scaling can be recovered with high precision when we model the \yxm\, scaling with a SBPL model, which leads to accurate predictions for the \ysz\,--\,\myx\, scaling at both cluster and galaxy group scales.
Bottom panels show the best-fit slopes for the \ysz\,--\,\myx\, scaling relation as a function of the \yx--\,mass proxy. 
}
\label{fig:Ysz_Myx_bestfits}
\end{figure*}

Our predictions for the three different models for \myx($M$)\, are presented in Figure~\ref{fig:Ysz_Myx_bestfits}. In the left panel, we show the dependence of the \ysz\, signal on the \yx\,--mass proxy, \myx, for quantities measured inside the true simulation aperture \rsim. 
We start by exploring the simplest model, which assumes that  \yx\, scales self-similarly with halo mass  (\textit{Assumption 1}). 
In Figure~\ref{fig:Ysz_Myx_bestfits}, this model generates the blue stars which mark the geometric means of the relation \ysz\,--\,\myx$_{,\rm self-similar}$. For comparison, we also include a dashed black line which marks the theoretical model for a self-similar dependence between \ysz\, and \mtot\,, i.e.,   \ysz\, $\propto M^{5/3}$. The normalization of this line was fixed to the best-fit value obtained from a simple power law fit on the highest mass clusters \mcrit\, $\geq 10^{14}$ \ms\, at $z=0$. For high mass clusters, the assumption of self-similarity in \yxm\, translates to an inferred \ysz\,--\myx\, relation that is very close to self-similar as well. 
 
We fit both a simple power law (SPL) and a smoothly broken power law (SBPL) model to the data points (blue stars) generated under \textit{Assumption 1}. In the bottom panel of Figure~\ref{fig:Ysz_Myx_bestfits}, we present the dependence of the slopes in the \ysz\,--\myx\, scaling relation on the \yx--mass proxy, \myx. 
As discussed in Section~\ref{sec:yxyszmassevolution}, SBPL models fit to \yxm\, and \yszm\, predict slopes at very high masses that are in good agreement ($\alpha_{\rm SBPL, 2}^{X} = 1.71$ and $\alpha_{\rm SBPL, 2}^{SZ} = 1.69$). However, \yxm\, predicts a shallower slope for galaxy and group scales ($\alpha_{\rm SBPL, 1}^{X} = 2.26$) than \yszm\, ($\alpha_{\rm SBPL, 1}^{SZ} = 2.42$). This difference in the slope predicted for lower mass objects in turn causes the \ysz\,--\myx$_{,\rm self-similar}$ scaling to have a slope slightly steeper than self-similar at small mass scales. The SBPL model fit for points generated under \textit{Assumption 1} predicts a self-similar slope at the highest mass end of $\alpha_{2, {\rm Assumption 1}} = 1.665$ and $\alpha_{1, {\rm Assumption 1}} = 1.750$ at the low mass end.

\textit{Assumption 2} modelled \yxm\, using a simple power law with a slope derived from the data (rather than assuming a self-similar slope). 
In Figure~\ref{fig:Ysz_Myx_bestfits}, we show the resulting \ysz\,--\myx$_{\rm SPL}$ data points using orange triangles. The inferred scaling relation is much steeper than self-similar. An SPL model fit to the data would predict that \ysz\, scales as \mtot$^{2.17}$. An SBPL model applied to the same \ysz\,--\myx$_{\rm SPL}$ data points under \textit{Assumption 2} predicts a very wide transition range between the two slopes, with $\alpha_{2, {\rm Assumption 2}} = 1.58$ and $\alpha_{1, {\rm Assumption 2}} = 2.35$.

We finally explore the scenario when the \yxm\, scaling relation is modelled using a SBPL model (\textit{Assumption 3}), which includes information about the break in the scaling relation and the mass range over which this break takes place. 
The predicted trend in the \ysz\,--\myx\, scaling is represented by red diamonds in Figure~\ref{fig:Ysz_Myx_bestfits}. Under  \textit{Assumption 3}, the scaling evolves similarly to the prediction of \textit{Assumption 2} for groups and galaxies below $10^{13}$\,\ms\, and it more closely matches the prediction of \textit{Assumption 1} above $10^{14}$\,\ms. A SBPL model applied to the data points generated under \textit{Assumption 3} predicts $\alpha_{2, {\rm Assumption 3}} = 1.70$ and $\alpha_{1, {\rm Assumption 3}} = 2.39$, with a break predicted to occur around $X_{\rm pivot} \simeq 3.79 \times 10^{13}$\,\ms.

Comparing to the results in Table~7 in the Appendix of Paper I, we find \textit{Assumption 1} that self-similarity holds for \yxm\, at all mass scales slightly biases low the slope measured at cluster scales by $\Delta \alpha = 0.021$. But more importantly, the self-similar assumption in \yxm\, completely hides away the break in the  true \yszm\, relation. Instead, it wrongly predicts an almost perfect self-similar scaling between \yszm\, at all mass scales and it overpredicts \ysz\, measurements at a fixed mass halo for low mass clusters, groups and galaxies. Assuming that \yxm\, follows a simple power law leads to an even stronger bias in the slopes predicted for high mass clusters, where the slope is too low by $\Delta \alpha = 0.108$ and the break in the \ysz\, -- \myx\, is shifted from $3.7 \times10^{13}$\,\ms\, to $5.7\times10^{13}$\,\ms. The exact details of the results under \textit{Assumption 2} depend on the mass distribution of the haloes used to measure the SPL \yxm\, best-fit. Nonetheless, smaller haloes will bias the SPL model to steeper slopes than self-similar for \yxm\, and this will result in overestimated values of \ysz\, at a fixed halo mass, especially for high mass clusters.

Finally, modelling the \yxm\, relation using a model that correctly captures the break in the scaling relation and the transition width for this break will lead to an inferred \ysz\, -- \myx\, scaling that is in very good agreement with the true \yszm\, scaling. This is the case for \textit{Assumption 3}, where we used a SBPL model for \yxm\, with the best-fits parameters from Table~6 in the Appendix of Paper I. The slopes predicted by the inferred \ysz\, -- \myx\, scaling ($\alpha_{2, {\rm Assumption 3}} = 1.70$ and $\alpha_{1, {\rm Assumption 3}} = 2.39$)  are in excellent agreement with the true slopes derived for \yszm\, in Paper I:
$\alpha_{2, \, Y_{\rm SZ}} = 1.687$ and $\alpha_{1, \, Y_{\rm SZ}} = 2.418$. Under this model, the break in the \yxm\, scaling was correctly modelled and this also translated in the inferred \ysz\, -- \myx\, scaling also having a break ($X_{\rm pivot} \simeq 3.79 \times 10^{13}$\,\ms) that is consistent with the break in the true \yszm\, scaling ($X_{\rm pivot} \simeq 3.7 \times 10^{13}$\,\ms).

 To summarize the results shown in Figure \ref{fig:Ysz_Myx_bestfits}, assuming that \yxm\, follows the self-similar prediction will almost entirely mask the true break in the \yszm\, scaling relation. In addition, a simple power law fit for  the \yxm\, scaling also biases the resulting  \ysz\, -- \myx\, relation, generating an artificially shallow slope for high mass clusters and over-predicting \ysz\, at a fixed halo mass. When using a SBPL model for \yxm\,, the inferred \yx\,--mass proxy provides highly precise mass estimates for the true halo mass, \mtot. The inferred \ysz\, -- \myx\, relation will, in this case, be in excellent agreement with the true \yszm\, scaling relation, matching the location of the break and the mass dependence of the slope  in \yszm\,  throughout the entire mass range considered by the study.

\subsection{Mass Bias from Modelling of X-ray Scalings}
\label{sec:massbiaslast}

\begin{figure*}
\centering
\includegraphics[width=0.7\textwidth]{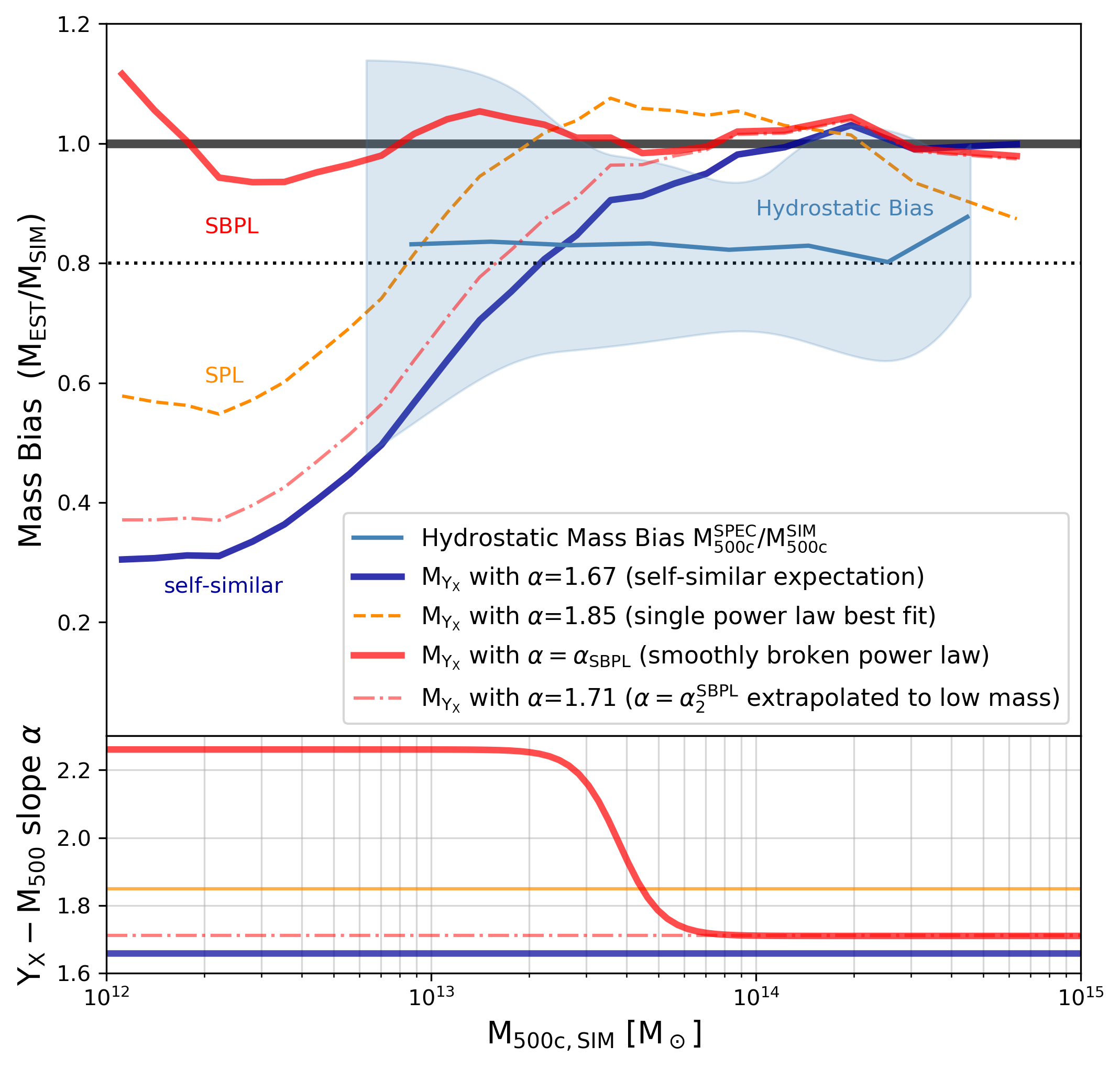}
\caption{Ratio between the estimated mass and the true simulation mass as a function of halo mass, \msimc. The light blue line and blue contours mark the median of the hydrostatic mass bias (\mspecc\,/\,\msimc), as well as 16\textsuperscript{th} and 84\textsuperscript{th} percentiles around that value. Hydrostatic mass bias introduces a significant uncertainty in the estimates of X-ray masses, with an average bias of ${\rm b}_{\rm X-RAY} = 1 -$ \mspecc\,/\msimc\, $= 0.170\pm0.004$ and large scatter, as showcased by the blue contours. 
The common assumption of self-similarity between X-ray observables (e.g., \yx) and mass (dark blue line) also introduces significant bias for galaxy groups \msim $\lesssim 10^{14}$ \ms\,, with estimated masses underestimating true masses more and more with decreasing halo mass. The dashed-dotted line in coral shows the small improvement achieved by assuming that the slope derived from the SBPL model at the high mass end ($\alpha_2$) can be extrapolated down to galaxy and group scales. Simple power law models (orange dashed line) try to find a compromise for all mass scales, but in return they underestimate masses for both galaxies (\mcrit $\lesssim 2\times 10^{13}$\ms\,) and the highest mass clusters (\mcrit $\gtrsim 3\times 10^{14}$\ms\,), and they overestimate galaxy group masses. The SBPL model (red line) introduced in this study achieves significantly better mass estimates for a wide range of mass scales, $10^{12} \leq$ \mcrit $\leq 10^{15}$ \ms. }
\label{fig:MbiasMyx}
\end{figure*}

One of the most attractive features of scaling relations between X-ray observables and cluster masses is that they offer a new method of inferring true halo masses. As discussed in Section~\ref{sec:massbias_yxysz}, X-ray mass estimates obtained from the inferred gas density and temperature profiles suffer from a significant level of bias ($b = 0.170^{+004}_{-004}$ above $10^{13}$\ms). In addition, there is a large degree of scatter around this median bias value, which increases the uncertainties associated with using hydrostatic mass estimates for cluster cosmology. In Figure~\ref{fig:MbiasMyx}, we compare the X-ray hydrostatic mass bias (blue contours) to the X-ray masses inferred from observations of the \yx\, parameter.

A common simplifying assumption is that the \yx--$M$ scaling relation follows the self-similar prediction: \yx $\propto M^{5/3}$. In this scenario, the only free parameter is the choice of normalization, A, for the \yx--$M$ scaling. This is chosen either from an absolute mass calibration based on usually a small number of high mass clusters or by relying on normalization values from the literature. In Figure~\ref{fig:MbiasMyx}, we choose the normalization ($A = 13.3$) of the self-similar line (dark blue) by fitting a simple power law with fixed slope ($\alpha = \alpha_{\rm self-similar} = 5/3$) to the individual data points with \msim\, $\geq 10^{14}$\ms\, at $z=0$ in TNG300. Compared to the hydrostatic bias, the self-similar \yx--mass proxy, \myx, is an almost unbiased mass estimate for the highest mass objects. Nonetheless, this is only true as long as the normalization is chosen appropriately for the distribution of halo masses in the sample. For all haloes with \mcrit\, $\lesssim 10^{14}$\ms\,, the self-similar assumption starts to break, leading to \myx values that are increasingly underestimating the true mass, \msim. Below \mcrit\, $\approx 3\times 10^{13}$\ms\,, \myx\, derived assuming self-similar \yx $\propto M^{5/3}$ introduces larger mass biases than using X-ray hydrostatic estimates, \mspec. Moreover, while the value of the hydrostatic mass bias is relatively independent of cluster mass in this mass range, the ratio \myx\,/\msim\, develops a steep gradient in the galaxy group regime, leading to significant errors. 

As shown in Section~\ref{sec:yxyszmassevolution}, we find that the \yx--$M$ scaling relation deviates slightly from the self-similar expectation for the highest mass clusters. The SBPL model gives a best-fitting parameter $\alpha_2 = 1.71^{+0.02}_{-0.02}$ for \yxsim--\msim, which is slightly steeper than the self-similar slope $\alpha = 1.67$. Assuming that \yx--$M$ follows a simple power law with an adjusted slope equal to the one derived for high mass clusters from the SBPL model, we obtain values of \myx\, (dash-dotted coral line in Figure~\ref{fig:MbiasMyx}) that yield a mild improvement compared to the self-similar expectation. The slightly steeper slope ($\alpha_2 = 1.71$) extends the regime where \myx\, generates a relatively small mass bias down to \mcrit\,$\approx 7 \times 10^{13}$\ms. However, this approximation comes with the downside of estimated masses \myx\, biased high by as much as $4\%$ between \mbox{\msim \, $\in [8\times 10^{13}$\ms--$2\times 10^{14}$\ms$]$}.

For comparison, we also include in Figure~\ref{fig:MbiasMyx} the ratio of \myx\,/\msim\, for the scenario where the \yx--$M$ relation is modelled using a simple power law (dashed orange line). 
Since this approach tries to accommodate the dependence of \yx\, on halo mass using only two free parameters, we are left with an inflexible model that severely underestimates the masses of both galaxies (\mcrit\, $\lesssim 2\times10^{13}\,$\ms) and the masses of the largest clusters (\mcrit\, $\gtrsim 3\times10^{14}\,$\ms), while significantly overestimating galaxy group masses.  

Lastly, we present \myx\, estimates from fitting a SBPL model to the \yx--$M$ relation. The SBPL model achieves very low mass bias for a wide range of mass scales, $10^{12}$ \ms\, $\leq$ \mcrit\, $\leq 10^{15}$\ms\, (red line in Figure~\ref{fig:MbiasMyx}). The mass bias for this model is $<5\%$ for clusters and massive galaxy groups, and it remains  $< 10\%$ even for the lower mass galaxies in our sample.

The bottom panel of Figure~\ref{fig:MbiasMyx} shows the \yx--$M$ slope as a function of halo mass, \msim, for each of the models described above. Both the SPL model (orange line) and the SBPL model (red line) predict slopes steeper than self-similar at all mass scales. The high-mass slope, $\alpha_2 = 1.71$, predicted by the SBPL model for clusters above \mcrit\,$\approx 7\times 10^{13}\,$\ms\, comes close to the self-similar expectation, $\alpha_{\rm self-similar} = 1.67$. 

In a nutshell, the SBPL model is able to produce mass estimates with very small bias for the full mass range of our sample by accurately capturing the presence of a break in the \yx--$M$ scaling around a pivot point \xp $=3.8\times 10^{13}\,$\ms\, and with a transition region extending between $2\times 10^{13}\,$\ms\, and  $7\times 10^{13}\,$\ms.
\section{Conclusions}
\label{sec:conclusions_letter}

In this paper, we use a sample of over 30,000 haloes from the TNG300 simulation of the IllustrisTNG project in order to explore the level of bias introduced by the hydrostatic mass bias, the X-ray mass bias, and the \yx\,--mass proxy.
In estimating X-ray observables for simulated haloes spanning a mass range \mcrit\, $\in [10^{12} - 2\times 10^{15}]$\,\ms, we produce synthetic X-ray images and derive cluster and galaxy group properties using methods that are consistent with observational techniques. We summarize our main results below:
\begin{enumerate}
\item We evaluate the level of mass bias ($b = 1 - M_{\rm EST}/M_{\rm TRUE}$) in \tng\,, under the assumption that haloes are in hydrostatic equilibrium. When we use mass-weighted profiles for the density, $\rho(r)$, and temperature, $T(r)$, of the hot gas, we find an average bias $b= 0.125 \pm 0.003$ for haloes with \mcrit\,$\geq 10^{13}$\,\ms\, and  $b=$ \val[0.099]{0.008}{0.026} for haloes with \mcrit\,$\geq 10^{14}$\,\ms. The hydrostatic mass bias is reduced significantly for relaxed clusters, which have an overall bias of $\sim 5\%$ above $10^{14}$\,\ms. 
In \tng, mass biases calculated using $\rho(r)$ and $T(r)$ profiles derived from a mock X-ray analysis increase to $b = 0.17 \pm 0.004$ above $10^{13}$\,\ms. X-ray mass estimates exhibit similar scatter as the hydrostatic mass estimates, but the scatter in X-ray masses increases at smaller halo masses. 
This work provides robust statistics for mass biases estimated from simulated haloes -- we explore the X-ray mass bias on a sample that includes more than 2,500 IllustrisTNG galaxy groups and clusters. 

\item \tng\, data indicates the presence of a break in the \yxm\, and \yszm\, scaling relations. 
In the companion paper, we introduced a smoothly broken power law model (SBPL, eqn.~\ref{eqn:eqnSBPLletter}) that accurately captures the break in the scaling relations, the transition region around the break, as well as the dependence of the best-fitting slope on the halo mass. In Section~\ref{sec:yxyszmassevolution}, we presented a detailed comparison between the breaks in \yxm\, and \yszm. In \tng\,, both scaling relations are consistent with the self-similar prediction for very high mass clusters. At galaxy and group scales, we find that \yx\, has a shallower slope than \ysz. 
The location of the break in \yxm\, shows a remarkable level of agreement with the break in \yszm\,, with both scalings predicting a break around $4\times 10^{13}$\,\ms.
However, once we account for the effect of X-ray mass bias on the \yxm\, scaling relation, the preferred pivot point shifts to $\sim 7\times 10^{13}$\,\ms. Our findings also suggest that \yx\, is more sensitive to differences in the dynamical state of clusters compared  to \ysz. While relaxed clusters are still following a self-similar \yszm\, scaling, their slopes in \yx\, are biased low.

 \item Next, we evaluate X-ray mass estimates using \yx\, as a mass proxy and explore how these different underlying assumptions for the \yx--mass proxy affect the inferred \yszm\, scaling relation. We model several common choices used in cluster cosmology, such as assuming that \yxm\, follows a self-similar relation at all mass scales or that \yx\, follows a simple power law in mass. In addition to these models, we also use a numerically inverted approximation of \myx from the best-fitting SBPL model for \yxm.
Our results indicate that the simplifying assumption that \yxm\, is self-similar at all mass scales will almost entirely mask the true break in \yszm\, and it will overestimate \ysz\, at galaxy and group scales. Simple power law fits for the \yxm\, scaling also bias the resulting \ysz\, -- \myx\, relation, predicting too shallow slopes for high mass haloes and overestimating \ysz\, at cluster mass scales. 

\item We show that calibrating X-ray masses using our smoothly broken power law prediction for the \yxm\, scaling relation results in a scaling relation between observed \ysz\, and estimated \myx\, that reproduces the true \yszm\, scaling relation with very good accuracy. Calibrations using the SBPL result can match the location of the break from self-similarity, as well as the mass dependence of the slope in \yszm. Moreover, \myx\, estimates calibrated with this method will lead to \ysz\, -- \myx\, predictions that are not biased by the inclusion of lower mass clusters or galaxy groups in the sample. Thus, this method can allow future SZ surveys to robustly measure the \yszm\, scaling relation irrespective of the mass cut or distribution of halo masses in their sample.

\item Finally, we present results for X-ray mass bias estimates over three orders of magnitude in halo mass (Figure~\ref{fig:MbiasMyx}). Our findings reiterate that self-similar or SPL assumptions for the \yxm\, scaling relation lead to non-negligible mass biases in the regime of low mass clusters, and increasingly severe biases at the group and galaxy mass scales. The smoothly broken power law model provides a  robust way to derive the \yx--mass proxy, significantly reducing the level of mass bias for clusters, groups, and galaxies.  
\end{enumerate}

\section*{Acknowledgments}
The authors would like to thank David Barnes for his contributions to \textsc{Mock-X} and helpful discussions prior to a career move. 
The work in this paper was supported by the NASA Earth and Space Science Fellowship (NESSF 80NSSC18K1111) awarded to Ana-Roxana Pop. LH was supported by NSF grant AST-1815978. RW is supported by the Natural Sciences and Engineering Research Council of Canada (NSERC), funding reference CITA 490888-16. MV acknowledges support through NASA ATP 19-ATP19-0019, 19-ATP19-0020, 19-ATP19-0167, and NSF grants AST-1814053, AST-1814259, AST-1909831, AST-2007355 and AST-2107724. D.\,Nelson acknowledges funding from the Deutsche Forschungsgemeinschaft (DFG) through an Emmy Noether Research Group (grant number NE 2441/1-1). PT acknowledges support from NSF grant AST-1909933, AST-2008490, and NASA ATP Grant 80NSSC20K0502. The computations were run on the Odyssey cluster supported by the FAS Division of Science, Research Computing Group at Harvard University. This research made use of several software packages, including \verb+NumPy+ \citep{Harris2020}, \verb+SciPy+ \citep{Virtanen2020}, and \verb+matplotlib+ \citep{Hunter2007}.

\section*{Data Availability}
Data from the IllustrisTNG simulations used in this work are publicly available at the website \href{https://www.tng-project.org}{https://www.tng-project.org} \citep{Nelson2019b}.

\bibliographystyle{mnras}
\bibliography{thesis}

\bsp	
\label{lastpage}
\end{document}